\documentstyle[12pt,epsf]{article}
\def\fnote#1#2{\begingroup\def\thefootnote{#1}\footnote{#2}\addtocounter
{footnote}{-1}\endgroup}
\def\inbar{\vrule height1.5ex width.4pt depth0pt}
\def\IB{\relax{\rm I\kern-.18em B}}
\def\IC{\relax\,\hbox{$\inbar\kern-.3em{\rm C}$}}
\def\ID{\relax{\rm I\kern-.18em D}}
\def\IE{\relax{\rm I\kern-.18em E}}
\def\IF{\relax{\rm I\kern-.18em F}}
\def\IG{\relax\,\hbox{$\inbar\kern-.3em{\rm G}$}}
\def\IH{\relax{\rm I\kern-.18em H}}
\def\II{\relax{\rm I\kern-.18em I}}
\def\IK{\relax{\rm I\kern-.18em K}}
\def\IL{\relax{\rm I\kern-.18em L}}
\def\IM{\relax{\rm I\kern-.18em M}}
\def\IN{\relax{\rm I\kern-.18em N}}
\def\IO{\relax\,\hbox{$\inbar\kern-.3em{\rm O}$}}
\def\IP{\relax{\rm I\kern-.18em P}}
\def\IQ{\relax\,\hbox{$\inbar\kern-.3em{\rm Q}$}}
\def\IR{\relax{\rm I\kern-.18em R}}
\def\IT{\relax{\rm I\kern-.18em T}}
\def\ZZ{\relax{\sf Z\kern-.4em Z}}
\def\a{\alpha}   \def\b{\beta}     
   \def\k{\kappa}  
     \def\si{\sigma}

   \def\cD{{\cal D}}
\def\cF{{\cal F}}   
   
 \def\cO{{\cal O}}

 \def\bq{{\bar q}}

\def\fnote#1#2{\begingroup\def\thefootnote{#1}\footnote{#2}\addtocounter
{footnote}{-1}\endgroup}
\def\beq{\begin{equation}}  \def\eeq{\end{equation}}
\def\bea{\begin{eqnarray}}  \def\eea{\end{eqnarray}}
 \def\lleq#1{\label{#1}\eeq}

\def\notin{\ \hbox{{$\in$}\kern-.51em\hbox{/}}}

\def\del{\partial}

\begin{document}
\hfill{BONN--TH--94--29}
\vskip 1.4truein
\noindent
\centerline{\large {\bf SCALING BEHAVIOR IN STRING THEORY}}

\vskip .8truein
\centerline{\sc R.Schimmrigk
                \fnote{\diamond}{netah@avzw02.physik.uni-bonn.de}
           }

\vskip .2truein
\centerline{\it Physikalisches Institut, Universit\"at Bonn}
\vskip .05truein
\centerline{\it Nussallee 12, 53115 Bonn}

\vskip 1.8truein
\centerline{\bf ABSTRACT}
\vskip .2truein

\noindent
In Calabi--Yau compactifications of the heterotic string there
exist quantities which are universal in the sense that they are
present in every Calabi--Yau string vacuum. It is shown that such
universal characteristics provide numerical information, in the form
of scaling exponents, about the space of ground states in string
theory.
The focus is on two physical quantities. The first is the Yukawa
coupling of a particular antigeneration, induced in four dimensions by
virtue of supersymmetry. The second is the partition function of the
topological sector of the theory, evaluated on the genus one worldsheet,
a quantity relevant for quantum mirror symmetry and threshold
corrections.
It is shown that both these quantities exhibit scaling behavior with
respect to a new scaling variable and that a scaling relation exists
between them as well.

\renewcommand\thepage{}
\newpage

\baselineskip=20pt
\parskip=.2truein
\parindent=20pt
\pagenumbering{arabic}

\vskip .3truein
\noindent
One of the outstanding questions in string theory concerns the structure
of the space of groundstates. Much insight has been gained over the past
years into some of its salient properties for (2,2)--supersymmetric vacua.
Most of these results, such as
the discovery of mirror symmetry \cite{ms}, and the computation of the moduli
dependence of Yukawa couplings \cite{cgop91}, focus on the geometry
associated to  the topological characteristics of the internal manifold
\fnote{1}{Or their field theoretic counterparts in the framework of
          Landau--Ginzburg theories or conformal field theories.}.
Relevant in this context is the generation--antigeneration structure of the
heterotic string, encoded in the two nontrivial Hodge numbers
$(h^{(1,1)},h^{(2,1)})$ of the internal Calabi--Yau manifold. These
topological numbers provide a first parametrization of the space of
string vacua and mirror symmetry, an operation that exchanges
these two numbers, provides a profound technique via which it is
possible to extract physical information about individual ground states.

What is still missing is a framework which leads to insight into the
more detailed structure of the space of string vacua and which will
allow, eventually, to address problems like the vacuum degeneracy.
In the absence of a physical `theory of moduli space'
what is needed is some concrete information about this space, which we
might hope to be encoded in some numerical characteristics.
Such numerical information should be useful in providing a guideline as to
what exactly it is that a future theory of moduli space has to explain.
It is the purpose of this Letter to provide such characteristics,
in the form of scaling exponents, thereby taking the first steps toward
a quantitative understanding of the moduli space of the supersymmetric
closed string.

The quantities which I will focus on in the following are based on
the existence of a universal structure which exists on every Calabi--Yau
manifold by virtue of the fact that they are projective, {\it i.e.} they are
described by equations in a particular type of compact, complex, K\"ahler
manifold, so--called projective spaces.  On such spaces there exists a natural
structure, the so--called hyperplane bundle, denoted by $L$, which can
be restricted to the Calabi--Yau manifold embedded in the ambient projective
space. This bundle is of interest for physics because its first Chern class
$c_1(L)$ leads to a universal massless mode, present in every
Calabi--Yau manifold, which parametrizes one of the antigenerations
observed in four dimensions. Moreover, on a three--complex dimensional
Calabi--Yau manifold, the degree of this line bundle, defined as
\beq
L^3 \equiv \int_M c_1^3(L),
\eeq
 affords an interpretation as a Yukawa coupling $\k_L$ of the
antigeneration associated to $L$. Thus $\k_L$ determines
the strength of the Yukawa coupling of the corresponding
four--dimensional mode. A second characteristic of any line bundle is the
number $h_L$ of its global functions. For every Calabi--Yau manifold
the line bundle $L$ thus leads to a pair of quantities $(\k_L,h_L)$
which can be viewed as coordinates on the space of string vacua.
The question arises whether these new coordinates allow us to draw
some type of phase diagram for the spaces of interest, {\it i.e.}
whether they provide a new sort of `structure diagram' for
string vacua.

A final quanitity of interest which is induced by the line bundle $L$ is
\beq
L\cdot c_2 \equiv \int_M c_1(L)\wedge c_2(M),
\lleq{c2}
where $c_2(M)$ is the second Chern class of the Calabi--Yau 3--fold.
The physical interpretation of this quantity has been uncovered
in ref. \cite{bcov93}, where it was shown that the generalized index
\beq
\cF =\frac{1}{2} \int \frac{d^2 \tau}{{\rm Im}~\tau}~
         Tr\left[(-1)^F F_L F_R q^{H_L} \bq^{H_R}\right],
\eeq
introduced in \cite{cfiv92}, contains information about the threshold
correction of the gauge couplings in string theory and that it
is the key for the understanding of quantum mirror symmetry at one loop.
Here the integral is over the fundamental domain of the moduli space of the
torus, $F_{L,R}$ denote the left and right fermion numbers and the trace is
over the Ramond sector for both the left-- and right--movers
\fnote{2}{The contribution of the ground states of the supersymmetric Ramond
     sector to $\cF$  has to be deleted in order for the integral to
converge.}.
In lowest order this index essentially reduces to
$\frac{1}{24} \int_{M} K\wedge c_2(M)$, where $K$ is the K\"ahler form of
the manifold. Hence the second Chern class evaluated on $L$, (\ref{c2}),
defines the universal contribution $\cF^{\uparrow}_L \equiv L\cdot c_2/24$
to the large volume limit of this partition function.

A class of Calabi--Yau spaces that is particularly amenable to an analysis
in terms of the variables $\k_L$, $\cF^{\uparrow}_L$, and $h_L$ is
furnished by hypersurfaces embedded in weighted projective space
$\IP_{(k_1,...,k_5)}$. The complete class of such manifolds, consisting of
7,555 configurations, has been
constructed in \cite{ks94,krsk92}. The natural candidate for
a line bundle on such spaces is the pullback of the weighted form
of the hyperplane bundle  $L \equiv \cO^{(k)}_{\IP_{(k_1,...,k_5)}}$
defined on the weighted ambient space, where  \cite{d93,bk93}
\beq
k=lcm\{~\{gcd(k_i,k_j)|~i,j=1,..,5; i\neq j\} \cup
\{k_i| k_i ~ {\rm does~not~divide}~ \sum_{i=1}^5 k_i \} \}.
\eeq
The restriction of $L$ from the weighted ambient space to the embedded
Calabi--Yau manifold $M$ induces an antigeneration which will also be
denoted by $L$. The Yukawa coupling $\k_L$ of this antigeneration leads,
in the large volume limit, to the expression
\beq
\k_L = \int_M (j^*(c_1( \cO^{(k)}_{\IP_{(k_1,...,k_5)}}))^3
      =\left(\frac{\sum_{i=1}^5 k_i}{\prod_{i=1}^5 k_i}\right)~k^3.
\eeq

The number of global functions of the bundles
$L=\cO^{(k)}_{\IP_{(k_1,...,k_5)}}$ can, finally, be obtained as
\beq
h_L = \frac{1}{k!} \frac{\del^k}{\del t^k}
 \left(\frac{\left(1-t^{\sum_{i=1}^5 k_i}\right)}{\prod_{i=1}^5
     \left(1-t^{k_i}\right)} \right)~\rule[-4mm]{.1mm}{10mm}_{~t=0}.
\eeq

In the following these variables will be used to explore the structure
of the moduli space of Calabi--Yau manifolds.  In general the two numbers
$\k_L$ and $h_L$ are independent, completely uncorrelated, quantities.
Hence in a diagram of these numbers we might expect a random distribution of
data points, with no obvious pattern. The actual computation, the results
of which are shown in Figure 1, however uncovers an unexpected simplicity.

\vskip .3truein
%size 1,0.9
\centerline{ \epsfbox{ 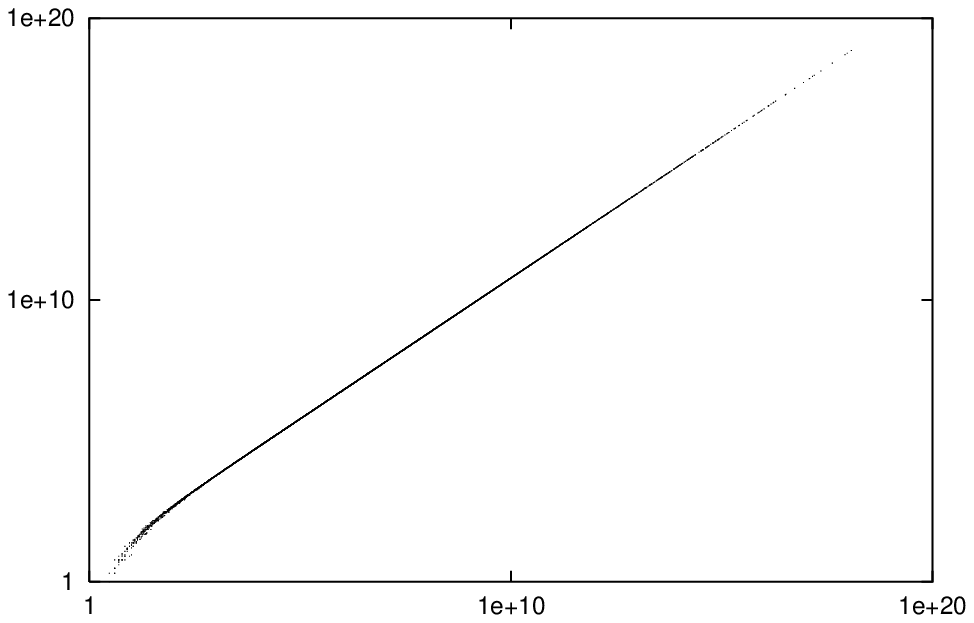}}

\vskip .2truein
%size 0.8,0.7
\centerline{{\bf Figure 1:} {\it The Yukawa couplings $\k_L$
                                 versus $h_L$.}}

The first thing to notice is the emergence of a well defined boundary
which is described by a simple relation: all varieties in the class of
Calabi--Yau 3--folds embedded in weighted $\IP_{(k_1,...,k_5)}$
satisfy the inequality
\fnote{3}{The weaker limit $\k_L < 6h_L$ follows from the relation of
          Riemann--Roch--Hirzebruch and the positivity of
          the second Chern class.}
\beq
\k_L \leq  2 \left(3h_L-8\right).
\lleq{yukasyl}
This is in contrast to mirror symmetry where the boundary of the
mirror diagram (see the first reference in \cite{ms}) does not yield to
such an easy description, even though it is also well defined.

What is rather unexpected is that as the dimensions $h_L$ increase for
the individual Calabi--Yau hypersurfaces the Yukawa couplings approach the
upper limit
\beq
\k_L^{as} \equiv 2(3h_L-8)
\eeq
very quickly, with very little scattering. Furthermore they do so according
to a power law.  This can be seen most clearly from Figure 2, which contains
the distances
\beq
\cD \equiv \frac{\k_L^{as}}{\k_L}-1
\eeq
of the Yukawa couplings from $\k_L^{as}$ as a function of $h_L$.

\vskip .3truein

%plotsize: 0.8,0.7
\centerline{ \epsfbox{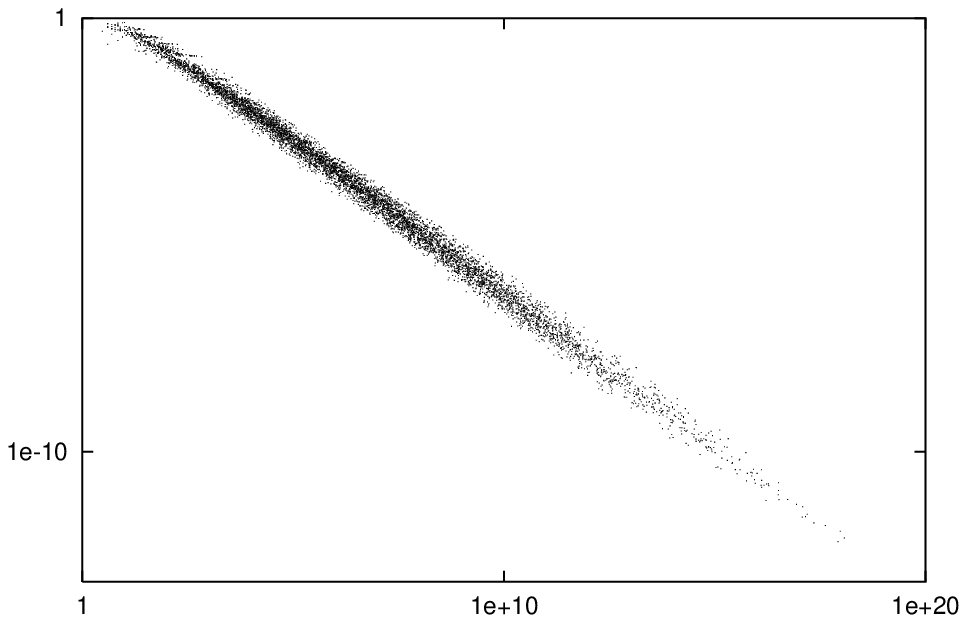}}

\vskip .1truein
\centerline{{\bf Figure 2:} {\it $\cD$ versus $h_L$.}}

The result shows that the distance variable $\cD$ exhibits the scaling behavior
\beq
\cD \sim A (h_L)^{-\alpha}~,
\lleq{discale}
where, approximately,
\beq
A=5 ~~~{\rm and}~~~\a = 0.7,
\eeq
via a least square fit. This leads to a behavior of the Yukawa couplings of
the form
\beq
\k_L  \sim  \frac{\k_L^{as}}{1+A (h_L)^{-\a} }
\lleq{yukpow}
and therefore the Yukawa coupling $\k_L$ behaves like an order parameter
with the dimension $h_L$ assuming the r\^{o}le of the scaling variable.

It should be noted that the Yukawa couplings $\k_L$ and the dimensions
$h_L$  take values over an enormously wide range of values, unprecedented
in Calabi--Yau physics, spanning 18 orders of magnitude. Thus they lead
to an enormous variation for the threshold corrections of these vacua.
Furthermore the power law
(\ref{yukpow}) for the coupling provides some measure of
simplicity of Calabi--Yau manifolds: the larger the value of the
scaling variable $h_L$, the more accurate the prediction of the Yukawa
coupling via (\ref{yukpow}). Since, very roughly, the magnitude
of the scaling variable is determined by the degree of complexity of the
singular sets of the varieties that need to be resolved, spaces with
more complicated resolution geometry are in fact simpler when it comes
to scaling. This phenomenon is rather different from the conventional
point of view which considers smooth projective hypersurfaces as the
simplest type of manifolds.  Finally, it is easy to see that the scaling
behavior (\ref{discale}), (\ref{yukpow}) is a characteristic of Calabi--Yau
manifolds, not of general projective 3--folds, not to mention general
3--folds: consider the hypersurfaces $\IP_4[d]$ of degree $d$
embedded in ordinary projective 4--space. For this (infinite) class of
manifolds one finds $\k_L =d, h_L=5, c_2\cdot L = d(d^2-5d+10).$

Over the last years a confluence of ideas has occured in physics and
mathematics toward a further physically important quantity. Wilson \cite{w93},
on the one hand, has reemphasized the importance of the linear form on the
Picard group, defined by the second Chern class $c_2$, for purposes of
classification, a fact first recognized by Wall.
Bershadsky, Cecotti, Ooguri, and Vafai, on the other have found that the
generalized index of \cite{cfiv92} reduces in the large volume limit
essentially to $\cF_L^{\uparrow}=c_2\cdot L/24$, and have shown that this
quantity arises in the 1--loop quantum mirror expansion
of the Ray--Singer torsion, as well as in the
threshold corrections of the gauge couplings. Thus the pair
$(\cF_L^{\uparrow}, \k_L)$ provides two coordinates on the moduli space
of Calabi--Yau manifolds with immediate physical meaning.
To investigate a possible relation between these quantities
is of particular interest since they arise in two different perturbation
expansions of the string -- the $\si$--model expansion at string tree level
and the string loop expansion.

Wilson \cite{w93} has observed that for the Yukawa coupling $\k_L$ and
$L\cdot c_2$ an inequality $L\cdot c_2 \leq 10 \k_L$ is obtained
for models whose Fujita index $\Delta$ is greater than two. The analysis
of the present class of weighted Calabi--Yau hypersurfaces shows that for
all but 16 spaces the Fujita index is always larger than two
\fnote{4}{None of these spaces is of particular interest from a physics
    perspective. }.
Hence Wilson's observation shows that except for those sixteen spaces the
above inequality does hold in this class, i.e.
\beq
\frac{\cF^{\uparrow}_L}{\k_L} \leq \frac{5}{12}.
\lleq{c2uplim}
No further information about the population below this upper limit is known.

The unexpected consequence of the analysis above of all Calabi--Yau
hypersurfaces in weighted projective 4--space
is that the vast majority of these vacua does not come even
close to the upper limit (\ref{c2uplim}) but instead takes values in a narrow
region far below that line.  For large values of $\cF^{\uparrow}_L$ the
Yukawa couplings lie more than ten orders of magnitude below the upper
limit provided by Wilson's observation.  Furthermore the results, as shown
in Figure 3, uncover a definite correlation between the Yukawa coupling and
$\cF_L^{\uparrow}$.

\vskip .3truein
%size 0.8,0.7
\centerline{ \epsfbox {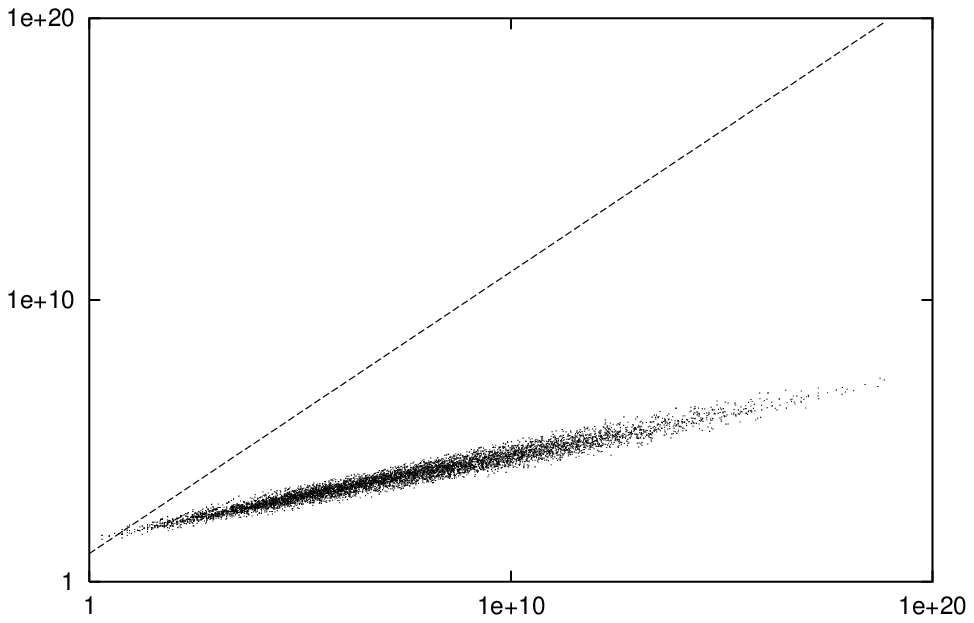}}

\vskip .2truein

\noindent
\centerline{{\bf Figure 3:}{\it~~ $24\cF^{\uparrow}_L$ versus $\k_L$.
              The dotted line represents the upper limit
              $y = \frac{5}{12} x$.}}

More precisely, a least square fit leads to the conclusion that
$\cF_L^{\uparrow}$ scales like
\beq
\cF^{\uparrow}_L \sim B ~\k_L^\beta
\eeq
with approximate values
\beq
B =1.5,~~~~\beta=0.29.
\eeq

To summarize, the results presented above suggest that the Yukawa coupling
$\k_L$ and the universal part $\cF_L^{\uparrow}$  of the free energy should
be interpreted as order parameters for which the number $h_L$ of global
functions behaves like a scaling variable. The scaling phenomena described
above provide specific power laws with a particular set of exponents that
any future physical theory of moduli space of the heterotic string will have
to explain. Thus our scaling laws furnish the first quantitative insight into
the structure of the string configuration space.

The present situation is not without historical precedent: it is
reminiscent of Bjorken's results many years ago when he argued that a certain
scaling behavior should manifest itself in deep inelastic electron--nucleon
scattering. Our scaling can in fact be presented in a form which is
very similar to the original data of these historical results.
Define a function $(h_L)^{-1} \k_L$ and consider its dependence
on the scaling variable $h_L$. The result, which is displayed in Figure 4,
 looks very much like the experimental results of the SLAC group for
$\nu W_2(\nu,q^2)$.

\vskip .2truein
%size 0.9,0.7
\centerline{ \epsfbox {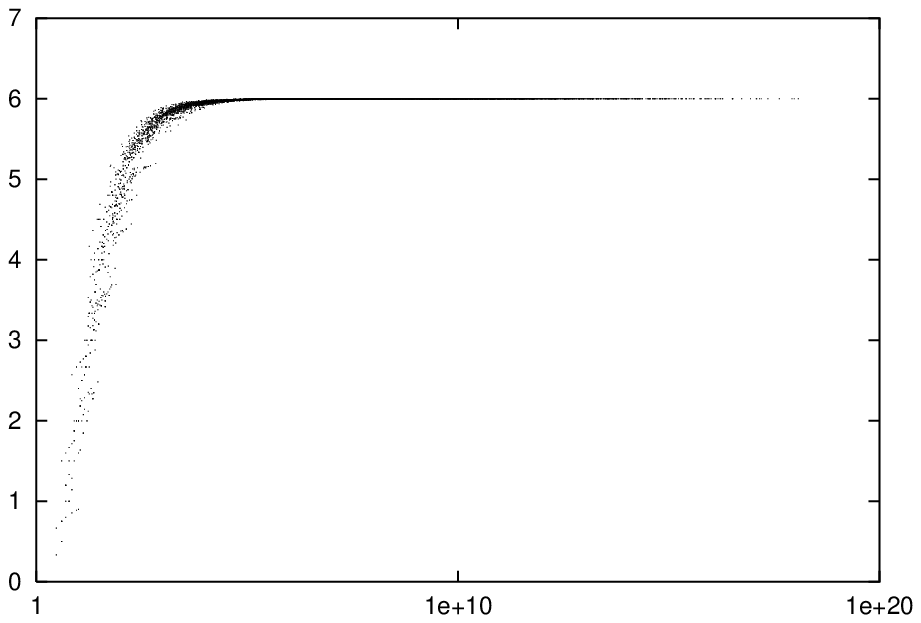}}

\vskip .2truein

\noindent
\centerline{{\bf Figure 4:}{\it~~ $(h_L)^{-1} \k_L$ versus $h_L$.}}

\noindent
What is needed, then, is the analog of Feynman's quark--parton picture, or
perhaps rather the analog of the QCD explanation of, say,
the momentum fraction of the proton carried by the quarks.

Tantalizing questions remain: An immediate one is
whether the exponents $\a,\b$ are universal or specific to our class of
Calabi--Yau hypersurfaces in weighted projective 4--space. There are two
natural avenues for further exploration. The simpler of the two is provided
by the toric framework of ref. \cite{b94} which has been shown \cite{cok94}
to include all Calabi--Yau hypersurfaces in weighted projective 4--space.
The second direction involves a natural generalization
of the framework of Calabi--Yau compactification, motivated by
mirror symmetry.  In \cite{rs93} a construction has been introduced which
provides an embedding of Calabi--Yau spaces into a larger class of a
special type of Fano manifolds, varieties whose  positive first Chern
class satisfies a particular quantization condition. It was shown in
\cite{rs93} that even though these special
Fano manifolds are not consistent ground states themselves they
do contain information about (2,2) supersymmetric string vacua.
It was furthermore demonstrated in ref. \cite{cod93} that such manifolds also
lead to the correct behavior of the Yukawa couplings and in ref. \cite{bb94}
that the framework of \cite{rs93} lends itself to a toric analysis,
generalizing the toric mirror construction of \cite{b94}. A natural question is
whether our scaling behavior persists in these more general settings as well.
Finally, the computations presented here are of lowest
order in perturbation theory and an interesting problem is to explore
the consequences of including higher order corrections.

\noindent
\section*{Acknowledgment}
It is a pleasure to thank Philip Candelas, Dimitrios Dais,
Xenia de la Ossa, Ed Derrick, Michael Flohr, Ariane Frey,
Jerry Hinnefeld, Vadim Kaplunovsky, Jack Morse, Werner Nahm, Steve Shore,
and especially Andreas Honecker, Monika Lynker and Katrin Wendland
for discussions.
I'm grateful to the Theory Group at the University of Texas at
Austin, the Department of Physics at Indiana University at South Bend,
and Simulated Realities Inc., Austin, TX for hospitality.

\end{document}